# Exogeology from Polluted White Dwarfs

Siyi Xu[1] and Amy Bonsor[2]

## ABSTRACT

It is difficult to study the interiors of terrestrial planets in the Solar System and the problem is magnified for distant exoplanets. However, sometimes nature is helpful. Some planetary bodies are torn to fragments and consumed by the strong gravity close to the descendants of Sun-like stars, white dwarfs. We can deduce the general composition of the planet when we observe the spectroscopic signature of the white dwarf. Most planetary fragments that fall into white dwarfs appear to be rocky with a variable fraction of associated ice and carbon. These white dwarf planetary systems provide a unique opportunity to study the geology of exoplanetary systems.

## INTRODUCTION

The chemical composition of a planet is a fundamental property, which is directly related to a planet's formation, evolution, and habitability. Both in the Solar System and other stellar systems, one of the best ways to investigate a planet's composition is to measure its mass and radius and derive a bulk density. A relatively high density implies a large metallic core, such as Mercury (5.4 g/cm$^3$) and K2-38b (11.0 g/cm$^3$) (Toledo-Padrón et al. 2020), while a relatively low density could mean the presence of volatile elements or an extended atmosphere. Therefore, whilst estimates of bulk density inform us about planet type, planet mass–radius relationships become increasingly degenerate when incorporating minor and trace elements (Dorn et al. 2015), so other techniques are needed to complete a picture of the compositions of planets. For example, in our own Solar System, we can measure meteorite compositions and then estimate a planet's bulk composition by assuming that meteorites are the remnants of planetary building blocks. In extrasolar planetary systems, white dwarfs offer an opportunity to break the degeneracy: they occasionally tear apart passing planetesimals, such as asteroids and comets, and even entire planets. We have an analog to meteorites in that the debris around white dwarfs were the building blocks of exoplanets (Jura and Young 2014).

[1] Gemini Observatory/NSF's NOIRLab
670 N. A'ohoku Place
Hilo, HI, 96720, USA
E-mail: siyi.xu@noirlab.edu

[2] University of Cambridge
Institute of Astronomy
Madingley Road
Cambridge CB3 0HA, UK
E-mail: abonsor@ast.cam.ac.uk

White dwarfs are the final evolutionary stages of stars that have a mass less than ten solar masses, which includes our Sun and most stars in our galaxy. In about six billion years, our Sun will use up its fuel and become a red giant star; it will expand and contract several times and expel about half of its mass. Afterwards, our Sun will collapse into a white dwarf, which has the size of the Earth but still retains about half of the original mass. White dwarfs have strong gravitational fields and rip apart any objects that come close to them. The critical distance at which such ripping occurs is called the "tidal radius", which depends upon the density ratio between the white dwarf and the passing object. Through various orbital perturbations, a planetesimal, or a planet, can cross the tidal radius of a white dwarf (Veras 2016). When this happens, the strong tidal force then breaks the object into fragments, and mutual collisions may grind the fragments down further into smaller pieces. This tidal disruption process would concurrently create a lot of dust and gas in orbit around the white dwarf. Eventually, this planetary debris will be accreted into the atmosphere of the white dwarf, creating a so-called "polluted" white dwarf.

A schematic view of white dwarf planetary systems is shown in FIGURE 1. Similar to the Solar System, there can be dust, gas, asteroids, comets, and planets orbiting around a white dwarf. One major difference is the scale: in a white dwarf planetary system, the size of the planets is comparable to or larger than that of the central star, while the tidal radius is comparable to the size of Saturn's rings. In addition, white dwarfs typically have a pure hydrogen or helium atmosphere, while the atmosphere of a Sun-like star has intrinsic metals (atomic number > 4) in its atmosphere. Therefore, when observing atmospheres of polluted white dwarfs, we are directly measuring the compositions of the extrasolar rocky planetesimals. In this sense, polluted white dwarfs can be thought of as cosmic mass spectrometers.

These kinds of observations are often performed with a high-resolution optical/ultraviolet spectrograph, which returns high-quality spectra of the white dwarf's atmosphere. White dwarf atmospheric models are then computed to match the observed spectra and to derive the abundances of each element. Here, we review the main results of extrasolar planet compositions from polluted white dwarf studies.

## TRACING THE FORMATION OF EXTRASOLAR PLANETESIMALS

We typically measure ~$10^{20}$ g of planetary debris in a polluted white dwarf's atmosphere, similar to the mass of a Solar System asteroid. The greatest accreted mass observed so far is in the white dwarf SDSS J0738, which has 7 x $10^{23}$ g of material, similar to the mass of the dwarf planet Ceres (Dufour et al. 2012). However, such mass estimates are snapshots in time, and, thus, represent a small fraction of the total mass accreted to a polluted white dwarf's atmosphere. These mass measurements, alongside the results from dynamical simulations (e.g., Veras 2016) show that we are likely measuring the compositions of extrasolar planetesimals, rather than an entire planet. A total of 23 different elements (i.e., Li, Be, C, N, O, Na, Mg, Al, Si, P, S, K, Ca, Sc, Ti, V, Cr, Mn, Fe, Co, Ni, Cu, and Sr) have been detected in polluted white dwarfs (e.g., Xu et al. 2014, Klein et al. 2021).

The detectability of an element depends on its absolute abundance and the white dwarf's properties, particularly its effective temperature and atmospheric composition. Detectability also depends upon the type of spectroscopy employed. High-resolution optical spectroscopy on the largest available optical telescopes is typically needed to detect the elements Ca, Si, Mg, Fe, and O, while ultraviolet (UV) spectroscopy is more sensitive to detecting trace elements C, N, and S. However, UV measurements require space telescopes, because UV photons are absorbed by Earth's atmosphere. The only currently available UV spectrograph with sufficient sensitivity and resolution to study polluted white dwarfs is the Cosmic Origins Spectrograph, which is onboard the *Hubble Space Telescope*.

FIGURE 2 shows a direct comparison between rocky objects in the Solar System and in extrasolar planetary systems from polluted white dwarf studies. Most planetary bodies that are accreted by white dwarfs appear to be dominated by the rock-forming elements Mg, Ca, and Fe, but may also include volatiles (e.g., C, N, O). This suggests that the material accreted by white dwarfs is like our own inner Solar System, which is full of rocky bodies (Mercury, Venus, Earth, the Moon, Mars, and the asteroids). Temperatures in the inner Solar System are hot and so volatiles tend not to take part in the planet formation process. Planetesimals that form at the hottest temperatures (closest to the star) can become rich in the very refractory species Ca, Al, and Ti, similar to the calcium–aluminum-rich inclusions (CAIs) found in meteorites in our own Solar System. Such compositions have been suggested to contribute to the high Ca, Al, and Ti abundances in the atmospheres of a handful of white dwarfs (e.g., Zuckerman et al. 2011; Harrison et al. 2018).

Some planetary debris in the atmospheres of white dwarfs, however, appears to be icy. Water molecules would dissociate in a white dwarf's atmosphere and, therefore, cannot be directly detected. So instead, we rely on the relative abundance of oxygen and other rock-forming elements to infer water content. Most oxygen will form "metal oxides", such as MgO, $SiO_2$, $FeO/Fe_2O_3$, and CaO, and after these have been accounted for, any excess accreted oxygen is likely to represent water. As shown in FIGURE 2, white dwarfs WD 1425, WD 1232, S1242, WDJ0738, and GD 61 have excess oxygen, and they are likely to have accreted ice-rich objects. In addition, WD 1425 also appears to have a significant amount of carbon and nitrogen, similar to a Kuiper Belt object (KBO) in our Solar System (Xu et al. 2017). These element ratios can also be used to assess oxygen fugacities in extrasolar planetesimals (Doyle et al. 2019). On the other hand, the mass fraction of carbon in extrasolar planetesimals ranges widely (e.g., Wilson et al. 2016), varying from 0.016% in PG 0843 to 9.6% in WD 1425, similar to the range observed in CV carbonaceous chondrites and Comet Halley. The range in volatile abundances shows that these extrasolar planetesimals have different thermal histories, indicating a remarkable diversity in the structure and evolution of planetary systems.

Exactly how volatiles are lost during planet formation is a subject of debate, even for our own Solar System. Certain elements can, however, act as tracers of the conditions under which volatiles are lost. Manganese and sodium are two examples and have been used to suggest that a planetary body in the atmosphere of white dwarf GD 362 lost its volatiles under oxidizing conditions and that these oxidizing conditions occurred only after the planetary system had lost its primordial nebula gas (Harrison et al. 2020).

## TRACING THE GEOLOGY OF EXTRASOLAR PLANETESIMALS

In addition to formation history, there are additional geological processes—such as differentiation into a core, mantle, and crust, followed by collisions—that can also change the distribution of chemical elements within a planet. For example, in Earth's early evolution, iron and iron-loving species (siderophile elements) sunk to the center of the molten planet to form a core, leaving behind a mantle that was rich in silicates. In our Solar System, core formation occurred even for planetesimals as small as 10 km in diameter. Some of these planetesimals were the building blocks for Earth, others became asteroids or meteorites. Iron meteorites are considered to be the metallic fraction of a larger differentiated object and give us clues as to how core formation occurred within the Solar System.

FIGURE 2 shows that several white dwarfs are rich in the planetary mantle elements Mg and Si compared to the core element Fe, whilst others are relatively rich in Fe. And just like iron meteorites, we see evidence that some white dwarfs have accreted fragments of planetesimals that were dominated by core material, whilst other accreted fragments were dominated by mantle materials, e.g., the broken bits of larger, differentiated planetary bodies. This is evident in the abundances of other trace species such as Ni, which follows Fe. FIGURE 3 shows the depressed Fe and Ni abundances of the planetary body in the atmosphere of the white dwarf GD 61 (Farihi et al. 2013), suggesting that it has accreted a mantle-rich planetesimal. The model, plotted in green, assumes Earth-like core–mantle differentiation and the loss of volatiles in chemical equilibrium. The depressed Fe, Ni, and Cr and the enhanced Al, Ti, Ca, and Na abundances of the planetary body in the atmosphere of white dwarf NLTT 43806 (Zuckerman et al. 2011), on the other hand, suggests that it has accreted a crust-rich planetesimal.

Most planetesimals will have experienced many collisions during their evolution through to the white dwarf phase. These collisions can, in some cases, dramatically alter the body, for example by chipping off mantle material or by the merging of cores from two planetary bodies. It is the fragments of these collisions that we observe in the white dwarf atmospheres. Detection of minor and trace elements in white dwarf atmospheres, such as Ba/Ca and Sr/Ca, can be used to search for evidence towards plate tectonics in extrasolar planetesimals (Jura et al. 2014). In summary, we can use the observations of planetary material in the atmospheres of white dwarfs to disentangle the complex collisional and geological history of extrasolar planetesimals.

## LOOKING FORWARD

Polluted white dwarfs provide an important means to measure the chemical compositions of extrasolar planetesimals. An ensemble of polluted white dwarfs will provide insights into an ensemble of extrasolar worlds. Thanks to the European Space Agency (ESA) space mission Gaia, hundreds of thousands of new white dwarfs have been discovered, increasing the current white dwarf sample by a factor of ten (Gentile Fusillo et al. 2019). Many new discoveries are on the horizon: the first transiting planet around a white dwarf was recently reported (Vanderburg et al. 2020). There are likely many more planets around white dwarfs yet to be discovered, and we will soon be able to measure their atmospheric compositions.

We can now look forward to studying the chemical compositions of numerous extrasolar planetesimals around white dwarfs via different techniques. Recent studies have found multiple gas species around polluted white dwarfs (e.g., Gänsicke et al. 2019). In the most extreme case, WD 1145 yielded ten different gas species (C, O, Mg, Si, Ca, Ti, Cr, Mn, Fe, Ni), which allows us to constrain the composition of not just the planetary debris in the white dwarf's atmosphere but also the gases that surround the star ("circumstellar" gas) (Fortin-Archambault et al. 2020). The *James Webb Space Telescope* is scheduled to launch in 2021, and it will have superior sensitivity such that the composition of the circumstellar dust can be measured (e.g., Reach et al. 2009). By comparing the polluted atmospheres of the white dwarfs to their circumstellar gas and dust, we will be able to further probe the full range of exoplanetary compositions, and have a basis on which to discuss many more aspects of exogeology.

## ACKNOWLEDGMENTS

We thank the referees Keith Putirka, Dimitri Veras, and Natalie Hinkel for useful comments that improve the manuscript. This work is partly supported by the international Gemini Observatory, a program of NSF's NOIRLab, which is managed by the Association of Universities for Research in Astronomy (AURA) under a cooperative agreement with the National Science Foundation, on behalf of the Gemini partnership of Argentina, Brazil, Canada, Chile, the Republic of Korea, and the United States of America. We thank the Royal Society for supporting this work, by a Royal Society Dorothy Hodgkin Research Fellowship, DH150130 and research grants, RG160509 and RGF\EA\180174.

## FIGURE CAPTIONS

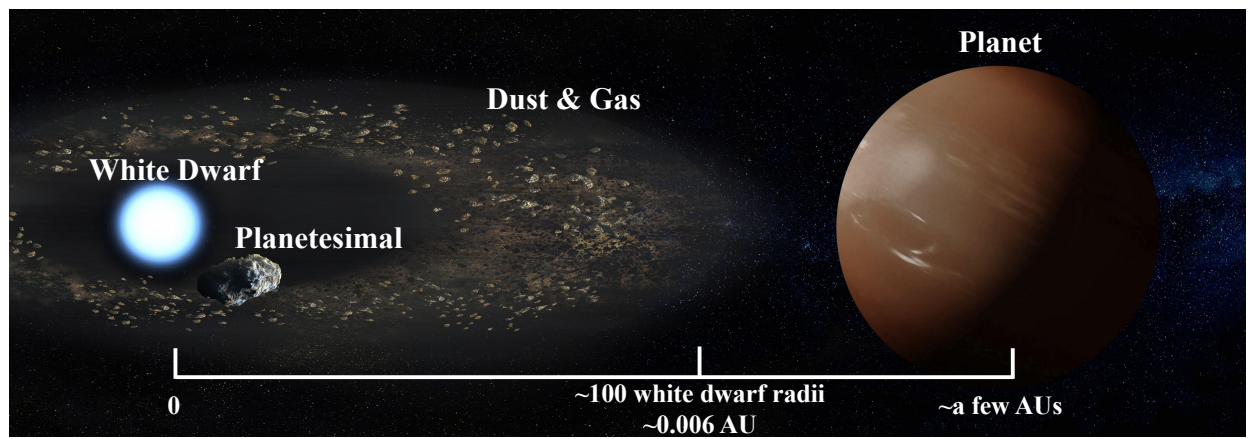

**FIGURE 1** A schematic view of white dwarf planetary systems. A planetesimal could be perturbed by a planet to enter into the tidal radius of the white dwarf and then disintegrate into dust and gas. Thereafter, the dust and gas would be accreted onto the surface of the white dwarf, creating a so-termed polluted white dwarf. AU = astronomical unit (distance from Earth to Sun used as a standard unit of distance). IMAGE COURTESY OF MICHEL H. GUAY.

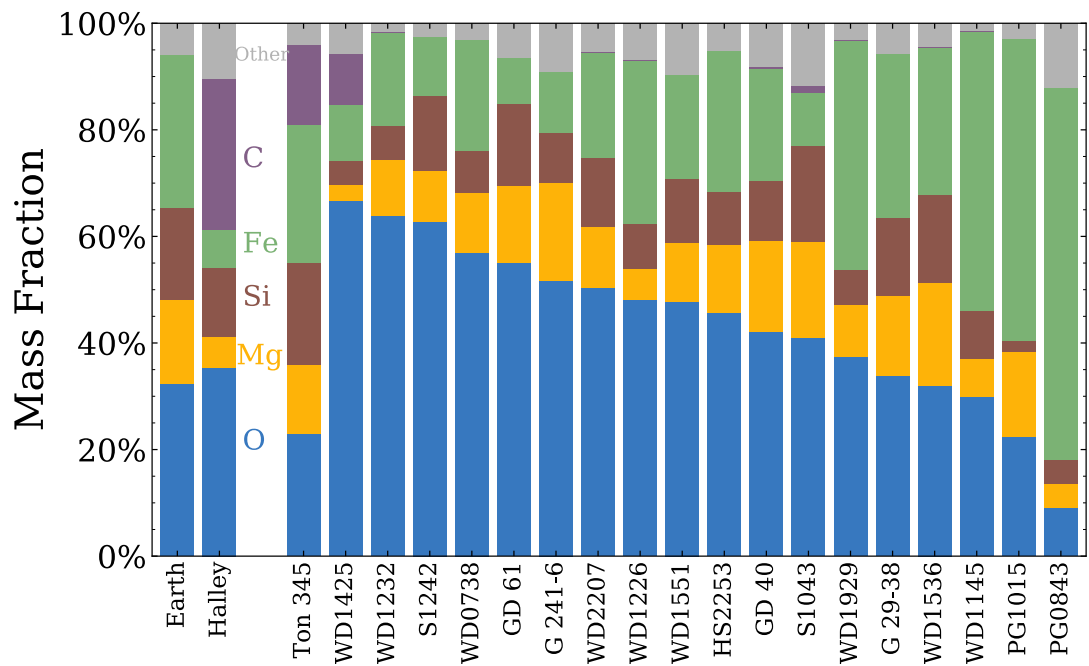

**FIGURE 2** Mass fractions of different elements—O, Mg, Si, Fe, C, and other—accreted onto polluted white dwarfs. The compositions of bulk Earth and comet Halley are also shown for comparison at left. Most white dwarfs (with their name designations given along horizontal axis) appear to have accreted rocky planetesimals.

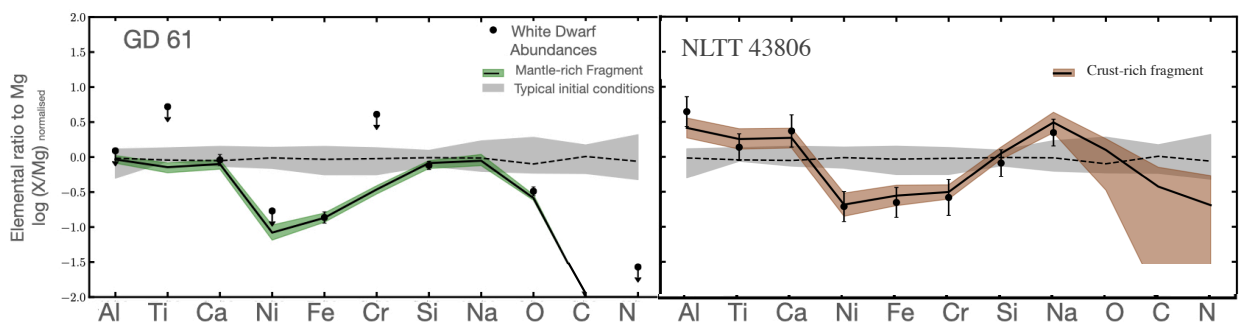

**FIGURE 3** (**LEFT**) The low Fe, Ni, and Cr abundances of planetary material in the atmosphere of white dwarf GD 61 suggests it has accreted a mantle-rich planetary body. (**RIGHT**) The low Fe, Ni, and Cr abundances plus the elevated abundances of Al, Ti, and Ca planetary material in the atmosphere of white dwarf NLTT 43806 suggests it has accreted a crust-rich planetary body. The black dots are measurements in the white dwarf atmospheres; the down arrows indicate upper limits. The abundances are normalized relative to solar abundances. The grey shaded regions indicate a plausible range of initial conditions for planet formation, based on compositions of nearby stars. The white dwarf compositions are very different from these initial conditions. AFTER ZUCKERMAN ET AL. (2011), FARIHI ET AL. (2013), AND HARRISON ET AL. (2018).